# LIGHT CURVE ANALYSIS OF W-UMa TYPE BINARY SYSTEMS RZ UMi, OQ UMa, LP UMa

M. Grozdanović and G. Djurašević

*Astronomical Observatory, Volgina 7, 11060 Belgrade 38, Serbia*

E–mail: *mgrozdanovic@aob.rs*

SUMMARY: This study presents new, high-quality, optical photometric observations of three W-UMa type contact binaries. The analysis of the corresponding light curves is made using Djurašević's inverse problem method. To explain the light-curve asymmetries and variations, we used the Roche model that involved regions containing spots on the components. The fundamental parameters of these systems were derived including the mass ratios determined using the $q$-search method. Hypotheses involving active surface regions, such as dark spots on the primary, and secondary components or bright spots in the neck region due to magnetic activity, or continuous mass transfer between components, were examined to explain the varying amplitudes of maxima and nearly equal minima depths in the light curves. Using the derived orbital and physical parameters, three-dimensional models were constructed for various orbital phases.

## 1. INTRODUCTION

A binary star system consists of two stars orbiting a common center of mass under the mutual gravitational influence of both components. It is estimated that between 50% and nearly 100% of all stars, whether in the vicinity of the Sun or further in the Galaxy, are part of a binary or, more generally, multiple systems (Abt 1983), with orbital periods ranging from 11 minutes to $10^6$ years (Podsiadlowski 2014). The majority of binary stars are systems where the components are sufficiently separated to lack physical interaction, allowing each star to evolve essentially as an isolated entity (Podsiadlowski 2014). However, a significant portion of stellar systems has orbital periods on the order of a few days, where the stars are so close to each other that mass transfer occurs from one component to the other, altering the structure of both stars and influencing their subsequent evolution.

Eclipsing binary stars are systems in which the brightness periodically changes in accordance with the orbital motion of the components. This occurs because the orbital plane of the system is almost edge-on to Earth, so that the stars orbiting their common center of mass, eclipse each other. As a result of these eclipses, photometric measurements register regular brightness variations, i.e. changes in magnitude over time. The light curve is obtained from these data and its analysis can provide a variety of information about the binary system: the type of eclipses, inclination of the orbital plane, radii and masses of the components, revolution period, ratio of effective temperatures, shape and orientation of the orbit, their separation, and, based on the derived parameters, a model of the system can be constructed to study the stellar evolution, as the components in binary systems have a common origin.

In such light curves, two types of minima can be observed, resulting from the primary or secondary eclipses. The relations of radii and inclinations primarily come from the duration and shape of the primary and secondary minima, whereas the temperature information is derived from their relative depths. The primary eclipse occurs when the brighter component is obscured by the less luminous component, while the secondary eclipse occurs when the less luminous component is eclipsed by the brighter one. If one star is significantly dimmer than the other, the secondary eclipse may not be easily noticeable to observers. The secondary eclipse occurs exactly midway between the primary eclipses if the stars' orbits are circular. In elliptical orbits, it is shifted depending on the shape and eccentricity of the orbit. The time between two successive primary minima is called the orbital period of the system and is not necessarily constant. If the stars are close to each other, the period can change due to mass transfer between the two stars. Small changes in the period can accumulate over time, causing the observed eclipse times to clearly vary from the predicted eclipse times.

Eclipsing binaries are usually also spectroscopic binaries, but the reverse is not always true. This is because eclipses are only likely in systems where the radius of one star is about 10% of their mutual separation, or if the inclination of the orbital plane is very close to 90° (Arbutina 2009), while there is no such limitation for changes in radial velocity.

Most eclipsing binary stars are close binary systems where stars can deviate from spherical shapes due to combined gravitational and centrifugal effects, which can lead to measurable brightness variations even when no eclipse occurs. These tidal distortions are manifested between eclipses as a subtle curvature in the light curve. Due to their proximity, the components of a close binary are mutually heated by radiation. This effect influences the temperature distribution across the stellar surfaces. On the hemispheres facing the companion star, the absorbed energy is redistributed and re-emitted. This reflection effect alters the shape of the light curve and it must be taken into account when modeling the synthetic light curve. It is also possible that stars in tight binary systems interact hydrodynamically and hydromagnetically, with gas flows and starspots on their surfaces, which can also affect the shape of the light curve in the form of slight asymmetries and changes in the curve over epochs.

Contact binaries (CB) of the EW type are characterized by continuous changes in brightness during eclipses and very strong gravitational effects that lead to ellipsoidal surfaces with varying gravity and brightness. The primary and secondary minima on the light curves have nearly equal amplitudes, and the periods are short, ranging from several hours to one day. The components come into contact when the more massive component, which evolves faster, fills its Roche lobe and transfers a large amount of material to its companion, which then fills its Roche lobe, forming a common convective envelope. The presence of a common envelope affects the dynamics and evolution of the system. Even for small mass ratios, the component temperatures are usually approximately equal (with temperature differences on the order of 100 K) (Lucy 1968) due to the presence of mass transfer from one component to the other, which also leads to changes in the orbital period and other physical characteristics of the system. The continuous brightness changes resulting from ellipsoidal variations, minima of nearly equal depth, and maxima that are not always symmetrical, are characteristic of their light curves. The difference in the amplitude of the light maxima, called the O'Connell effect (O'Connell 1951), is caused by inhomogeneity in the distribution of surface brightness on one or both stars and is associated with dark magnetic spots. The presence of spots on their surfaces is common and can lead to variations in the light curves, from which information about the magnetic activity and differential rotation of the components can be derived.

Two subclasses can be distinguished: the A-type and W-type systems. A-type is found among more massive stars of earlier spectral types from A to F, where, during the primary (deeper) minimum, the more massive primary component is eclipsed. The smaller star may have a slightly lower surface temperature, but this is not always the case (Alton and Stępień 2021). Among the less massive systems of the W subclass are stars of later spectral types from G to K, where the secondary, less massive component is eclipsed during the primary minimum. It is often the case that this less massive secondary component has a higher effective temperature than the primary. With their specific properties, the W UMa-type contact binary stars form a special group of objects that are easy to detect and identify due to the large amplitude of light variations reaching up to one magnitude and short orbital periods, making short-term monitoring sufficient.

## 2. OBSERVATIONS

For this study, three CB systems were observed at the Astronomical Station Vidojevica in Serbia during spring 2024. The CCD photometric observations of the systems RZ UMi and OQ UMa were conducted using the "Nedeljković" reflector telescope with a 0.6 m primary mirror. The FLI ProLine PL23042 CCD camera was mounted at the Cassegrain focus with a focal length of 6000 mm ($f/10$), providing a field of view of $17.5 \times 17.5$ arcminutes. Both stars were observed for one night, covering the entire orbital period. Observations were carried out using the wide Bessell B, V, Rc and Ic filters, closely aligned with the classical Johnson-Cousins system (Bessell 1995). The exposure times for the RZ UMi system were 120, 45, 30, and 30 seconds for the B, V, R, and I filters, while for the OQ UMa system, the exposure times were 90, 45, 45, and 45 seconds, respectively. Comparison and control stars were selected from the fields of view containing the systems chosen as to closely

Table 1: Coordinates, magnitudes, and periods of the observed stars

| Star | Name | R.A.$_{2000}$ [h:m:s] | Dec$_{2000}$ [°:′:″] | G[mag] | Period [d] |
|---|---|---|---|---|---|
| Star | RZ UMi | 14:54:26.02 | +86:43:37.41 | 11.6298 | 0.3373 |
| Comparison | GSC 04642-00752 | 14:49:12.10 | +86:45:43.44 | 13.6392 | - |
| Control | GSC04642-00777 | 14:48:47.58 | +86:43:41.29 | 13.8727 | - |
| Star | OQ UMa | 13:57:22.36 | +56:26:06.93 | 13.0022 | 0.2833 |
| Comparison | Gaia DR3 1657776085411370368 | 13:57:07.86 | +56:25:20.11 | 13.3334 | - |
| Control | Gaia DR3 1657776257210062592 | 13:57:12.66 | +56:27:13.99 | 14.7702 | - |
| Star | LP UMa | 10:33:57.79 | +58:52:15.55 | 12.4782 | 0.3090 |
| Comparison | GSC 03822-00070 | 10:33:48.79 | +58:52:30.83 | 13.0574 | - |
| Control | Gaia DR3 855120605585788160 | 10:34:22.49 | +58:52:45.39 | 13.9315 | - |

match the position and brightness of the binary systems.

The LP UMa system was observed using the "Milanković" reflector telescope with a 1.4 m primary mirror. The telescope uses an alt-azimuthal mount and a Ritchey-Chrétien optical system, which provides a field of view of 30 arcminutes without significant aberrations (Vince 2021). The effective focal length at all outputs is 11200 mm ($f/8$). A focal reducer is attached at one of the Nasmyth exits, reducing the focal length to 7132 mm. The iKonL CCD camera, which provides a field of view of $13.3 \times 13.3$ arcminutes, is mounted at this exit. The LP UMa system was observed for one night using the B, V, Rc, and Ic filters according to Bessell's specifications, with exposure times adjusted during the night as to maintain a linear camera range as the airmass changed with the increasing altitude of the star.

Twilight flat fields were obtained for each filter, and dark and bias frames were also taken throughout the runs. The frames were combined into a single master bias, master dark, and master flat frame, respectively. The standard procedure was used for the reduction of the photometric data (debiasing, dark frame subtraction, and flat-fielding) using the PixInsight software (Keller 2018).

Aperture photometry was performed using the AstroImageJ software package (Collins et al. 2017) with comparison and check stars as indicated in Table 1[1].

## 3. LIGHT CURVE SOLUTIONS

To analyze asymmetric light curves deformed by the presence of spotted areas on the components, we used Djurasevic (1992a) program generalized to the case of an overcontact configuration (Djurasevic et al. 1998) which is characteristic for the W UMa-type systems. The program is based on the Roche model and the principles arising from the paper by (Wilson and Devinney 1971). The light curve analysis was performed applying the inverse-problem method (Djurasevic 1992b) based on the Marquardt (1963) algorithm. According to this method, the stellar size in the model is described by the filling factors for the critical Roche lobes $F_{1,2}$ of the primary, and secondary component, respectively, which tell us to what degree the stars in the system fill their corresponding critical lobes.

For synchronous rotation of the components, the filling factors are expressed as the ratio of the stellar polar radii, $R_{1,2}$, to the corresponding polar radii of the critical Roche lobes, $F_{1,2} = R_{1,2}/R_{\text{Roche}1,2}$. The dimensionless polar radii, expressed relative to the orbital separation, were obtained from the solution of the light curve. The absolute radii of the components were then derived by scaling with the orbital separation, which was computed from the total mass of the system and orbital period. Since the mass of the primary component was estimated using Cox (2000) for the main-sequence stars, and the mass ratio was found photometrically, the absolute radii of both stars follows directly from the light-curve solution. This procedure was systematically applied to all three systems studied in this paper. In the case of an overcontact configuration, the potential $\Omega_{1,2}$ characterizing the common photosphere is derived with the filling factor of the critical Roche lobe $F_1 > 1$ of the primary, while the factor $F_2$ may be excluded from further consideration. The degree of overcontact is defined in the classical way (Lucy and Wilson 1979) as: $f_{\text{over}}[\%] = 100 \cdot (\Omega_{1,2} - \Omega_i)/(\Omega_o - \Omega_i)$, where $\Omega_{1,2}$, $\Omega_i$, and $\Omega_o$ are potentials of the common photosphere and of the inner and outer contact surfaces, respectively. To achieve more reliable estimates of the model parameters in the light-curve analysis code, we applied a fairly dense coordinate grid, having $72 \times 144 = 10368$ elementary cells per star. The intensity and angular distribution of radiation of elementary cells are determined by the stellar effective temperature, limb-darkening, gravity-darkening, and by the effect of reflection in the system. The presence of spotted areas (dark or bright) enables us to explain the asymmetries and the light-curve anomalies. In our code, these active regions are approximated by circular spots, characterized by the temperature contrast of the spot with respect to the surrounding photosphere ($A_{cs} = T_{cs}/T_s$), by the angular dimension (radius) of the spot ($\theta_s$), and by the longitude ($\lambda_s$) and latitude ($\phi_s$) of the spot center. The longitude ($\lambda_s$) is measured clockwise (as viewed from the direction of the $+Z$-axis) from the $+X$-axis (the line connecting star centers) in the range $0° - 360°$. The latitude ($\phi_s$) is measured from $0°$ at the stellar equator (the orbital plane) to $+90°$ towards the "north" ($+Z$) and $-90°$ towards the "south" ($-Z$) pole.

---
[1] The stellar magnitudes presented in the table are sourced from the GAIA Data Release 3 (DR3) catalog, based on high-precision photometric measurements performed by the GAIA space mission.

The optimal model parameters are obtained through the minimization of:

$$S = \sum_{i=1}^{n}(O_i - C_i)^2$$

where $O - C$ is the residual between the observed (LCO) and synthetic (LCC) light curves for a given orbital phase. The standard deviation of the observations is calculated with:

$$\sigma = \sqrt{\frac{\sum_{i=1}^{n}(O_i - C_i)^2}{(n-1)}}.$$

The minimization of $S$ is done in an iterative cycle of corrections of the model parameters. In this way, the inverse problem method gives us estimates of the system parameters and their standard errors. The values of the limb-darkening coefficients were derived from the stellar effective temperature and surface gravity, according to the given spectral type, using the polynomial proposed by (Diaz-Cordoves et al. 1995).

During the optimization process, according to the temperature changes, we have an automatic re-computation of the limb-darkening. Following Lucy (1967), Ruciński (1969) and Rafert and Twigg (1980), the gravity-darkening coefficients of the stars, $\beta_{p,s}$, and their albedos, $A_{p,s}$, were set at values of 0.08 and 0.5, respectively, appropriate for stars with convective envelopes. The present analysis yields $F_p > 1$ for the filling coefficient in the critical Roche lobe, i.e., the overcontact configuration. Tidal effects are expected to contribute to synchronization of the rotational and orbital periods. Therefore, in the inverse problem, we adopted $f_{p,s} = \omega_{p,s}/\omega_K = 1.0$ for the nonsynchronous rotation coefficients, where $f_{p,s}$ is the ratio of the angular rotation rate ($\omega_{p,s}$) to the Keplerian ($\omega_K$) orbital revolution rate. The surface gravities can be derived very accurately from masses and radii of CB stars by solving the inverse problem of the light-curve analysis.

The parameters derived from the light-curve analysis are listed in tables for each star. The errors of the parameter estimates arise from the nonlinear least-squares method, on which the inverse-problem method is based. The indices (B, V, R, I) denote the B, V, R, and I band observations, respectively. In the same tables, the spot characteristics are also given. The determination of these parameters is based on a simultaneous fitting of the available light curves for all given photometric bands for different epochs of observations with the same set of basic system parameters.

### 3.1. RZ UMi

The eclipsing binary system RZ UMi was discovered by Goranskij (Goranskij 1982), who classified it as a member of the W UMa-type class, with a period of $P = 0.3373$ days (Hoffman et al. 2009). Additional observations were made by Van Cauteren (van Cauteren et al. 2006) using a $0.4\,m$ Newtonian telescope and V filter according to Bessell's specifications (Bessell 1995). The system parameters were derived using the Wilson and Devinney code (Wilson and Devinney 1971). A rough period-color relationship for the W UMa stars (Eggen 1967) was used to assume an average temperature of $T_1 = 5500$ K for the primary star. Gravitational darkening values ($g_{p,s} = 0.32$) and bolometric albedo values ($A_{p,s} = 0.5$), typical of contact binary systems, were assumed. The best fit was then achieved with free parameters for the temperature of the secondary component ($T_2$), the inclination ($i$), the surface potential and the relative monochromatic brightness of the primary star, assuming different values of the mass ratio $q$. These calculations showed that any mass ratio between 0.8 and 2.2 fit the data almost equally well. The secondary star temperature remained nearly constant at $T_2 = 5395 \pm 25$ K throughout this range, while the inclination increased from $i = 80.3°$ for $q = 0.8$ to $i = 83.3°$ for $q = 2.2$.

With new and higher-quality observational data from this study, a more detailed analysis of this system was performed. The initial observational data, provided through the time of observation expressed in heliocentric Julian days and differential magnitudes, are shown in Table 2.

**Table 2**: Section of the light curve data around the secondary minimum for the RZ UMi system in the I filter

| HJD | Phase | $\Delta m$ | Error |
|---|---|---|---|
| 2460390.3265 | 0.4683 | -0.8746 | 0.0066 |
| 2460390.3293 | 0.4766 | -0.8135 | 0.0065 |
| 2460390.3321 | 0.4850 | -0.7700 | 0.0068 |
| 2460390.3349 | 0.4933 | -0.7683 | 0.0066 |
| 2460390.3377 | 0.5016 | -0.7649 | 0.0065 |
| 2460390.3405 | 0.5099 | -0.7575 | 0.0062 |
| 2460390.3433 | 0.5182 | -0.7773 | 0.0069 |
| 2460390.3461 | 0.5265 | -0.8279 | 0.0065 |
| 2460390.3489 | 0.5348 | -0.8798 | 0.0066 |
| 2460390.3518 | 0.5432 | -0.9432 | 0.0065 |
| 2460390.3546 | 0.5515 | -1.0019 | 0.0064 |
| 2460390.3574 | 0.5598 | -1.0479 | 0.0065 |

By knowing the system's orbital period and light curve, through the epoch of the primary minimum, these observational data can be easily transformed using the standard method into a form where the time scale is converted into orbital phases. This allows for the construction of a light curve as a function of differential magnitude and phase. The resulting light curve is of the EW type, showing the asymmetry and different amplitudes of the maxima, as well as the widths of the primary and secondary minima. Using the method of Kwee and van Woerden (Kwee and van Woerden 1956) , the times of the primary and secondary minima were calculated as follows:

$$I_{min} = 2460390.5059 \pm 0.0001,$$

$$II_{min} = 2460390.3376 \pm 0.0001,$$

The new obtained ephemeris is:

$$I_{min}[HJD] = 2460390^d.5059 + 0.3373 \times E.$$

With the light curve obtained, it is possible to perform an analysis, incorporating spots on the components to explain the differences in maximum heights and asymmetry. The analysis was performed using the inverse problem method for light curves in the B, V, Rc, and Ic bands separately.

In the absence of spectroscopic observations that would provide values for radial velocities, the analysis begins by determining the mass ratio solely from photometric data using a $q$-search method. The minimum value of $\sum(O-C)^2$ is sought by varying the parameter $q$. For this system, the obtained value is $q = m_2/m_1 = 0.570$ (Fig. 1). Furthermore, the change in orbital inclination with respect to variation in the mass ratio was plotted, resulting in an inclination value $i \approx 85.2 \pm 0.6$, corresponding to a mass ratio $q = 0.570$. Assuming this mass ratio value, two hypotheses were examined. The first hypothesis assumes a cool spot on the secondary component, while the second assumes a cool spot on the primary component. The temperature of the primary component was fixed at $T_1 = 5500$ K, while the model optimization resulted in a slightly higher temperature for the secondary component: $T_2 = 5650 \pm 20$ K. The system is in contact configuration with fill-out factor of $f_{over} \approx 9\%$.

The results show an excellent agreement with the solutions under these working hypotheses regarding the presence of a dark spot on either the primary or secondary component. The final sum of squared deviations between the observed and synthetic model light curves is small, indicating a good fit. The left part of Table 3 presents the parameters obtained and their determination errors for the hypothesis with a cool spot on the secondary component, while the right part provides the same for the hypothesis with a cool spot on the more massive primary component, both derived using the least squares method for each hypothesis individually. Based on the results obtained, Figs. 2 and 3 show the observed (marked with dots) and final synthetic light curves (marked with a solid line), overlaid to visually assess the quality of the fit. The final $(O-C)$ deviations between the observed and optimal synthetic light curves, obtained using the optimal model parameters by solving the inverse problem, are within the measurement accuracy. The deviations are shown (marked with dots) below the light curves with the label $O-C$ for each filter separately. Based on the obtained results, RZ UMi can be classified as an A-subtype system, as the more massive component is eclipsed in the primary minimum. On the right-hand side of Figs. 2 and 3, a 3D model of the system is shown, as it would appear to an observer at orbital phases 0.05, 0.25, 0.60, 0.82, 0.90, and 0.95.

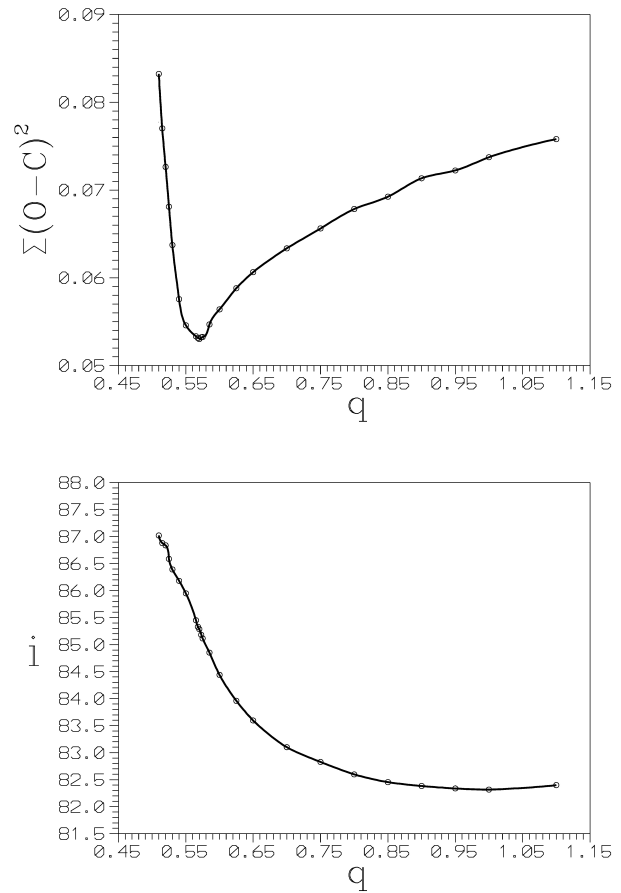

Fig. 1: Searching for the mass ratio and inclination using the $q$-search method. Each point on the upper graph represents the quality of the light curve fit, $\sum_i(O-C)^2$ for the assumed mass ratio of the components.

### 3.2. OQ UMa

OQ UMa was discovered as a variable star during a sky survey around the galaxy M101 (Romano 1979). New photometric observations of the system were conducted on the night of April 8, 2024. Until now, no quantitative light curve analysis of this system has been performed, so, with the new observational data, we proceed to construct and analyze the light curve. Table 4 presents a portion of the observational data obtained by differential photometry, which is used to construct the light curve.

**Table 3**: The results of the simultaneous analysis of the BVRI light curves of the close binary system RZ UMi, obtained by solving the inverse problem for the Roche model with a cool spot on the less massive, secondary component (left) and with a cool spot on the more massive, primary component (right).

| **Cool spot on secondary** | | **Cool spot on primary** | |
|---|---|---|---|
| Quantity | | Quantity | |
| system | RZ UMi | system | RZ UMi |
| $n(B+V+R+I)$ | 649 | $n(B+V+R+I)$ | 649 |
| $\Sigma(O-C)^2$ | 0.0509 | $\Sigma(O-C)^2$ | 0.0516 |
| $\sigma_{rms}$ | 0.0089 | $\sigma_{rms}$ | 0.0089 |
| $q=m_s/m_p$ | 0.570 | $q=m_s/m_p$ | 0.570 |
| $T_p[K]$ | 5500 | $T_p[K]$ | 5500 |
| $A_{p,s}$ | 0.5 | $A_{p,s}$ | 0.5 |
| $\beta_{p,s}$ | 0.08 | $\beta_{p,s}$ | 0.08 |
| $f_p=f_s$ | 1.0 | $f_p=f_s$ | 1.0 |
| $A_{cs}=T_{cs}/T_s$ | $0.96 \pm 0.02$ | $A_{cs}=T_{cs}/T_s$ | $0.96 \pm 0.02$ |
| $\theta_{cs}$ | $48.7 \pm 1.3°$ | $\theta_{cs}$ | $48.7 \pm 1.3°$ |
| $\lambda_{cs}$ | $311.7 \pm 4.2°$ | $\lambda_{cs}$ | $300.2 \pm 4.2°$ |
| $\varphi_{cs}$ | $71.6 \pm 1.5°$ | $\varphi_{cs}$ | $74.2 \pm 1.5°$ |
| $T_s[K]$ | $5650 \pm 20$ | $T_s[K]$ | $5620 \pm 20$ |
| $F_p$ | $1.012 \pm 0.002$ | $F_p$ | $1.013 \pm 0.002$ |
| $i$ | $85.2 \pm 0.6°$ | $i$ | $85.3 \pm 0.6°$ |
| $l_3/(l_1+l_2+l_3)_B$ | $0.000 \pm 0.002$ | $l_3/(l_1+l_2+l_3)_B$ | $0.001 \pm 0.002$ |
| $l_3/(l_1+l_2+l_3)_V$ | $0.000 \pm 0.002$ | $l_3/(l_1+l_2+l_3)_V$ | $0.003 \pm 0.002$ |
| $l_3/(l_1+l_2+l_3)_R$ | $0.000 \pm 0.002$ | $l_3/(l_1+l_2+l_3)_R$ | $0.004 \pm 0.002$ |
| $l_3/(l_1+l_2+l_3)_I$ | $0.012 \pm 0.002$ | $l_3/(l_1+l_2+l_3)_I$ | $0.016 \pm 0.002$ |
| $\Omega_{p,s}$ | 2.9768 | $\Omega_{p,s}$ | 2.9760 |
| $\Omega_{in}$ | 3.0081 | $\Omega_{in}$ | 3.0081 |
| $\Omega_{out}$ | 2.6725 | $\Omega_{out}$ | 2.6725 |
| $f_{over}[\%]$ | 9.30 | $f_{over}[\%]$ | 9.56 |
| $R_p[D=1]$ | 0.408 | $R_p[D=1]$ | 0.408 |
| $R_s[D=1]$ | 0.315 | $R_s[D=1]$ | 0.315 |
| $L_p/(L_p+L_s)(B;V;R;I)$ | 0.600; 0.602; 0.603; 0.605 | $L_p/(L_p+L_s)(B;V;R;I)$ | 0.585; 0.588; 0.591; 0.593 |
| $m_p[M_\odot]$ | $0.92 \pm 0.02$ | $m_p[M_\odot]$ | $0.92 \pm 0.02$ |
| $m_s[M_\odot]$ | $0.52 \pm 0.02$ | $m_s[M_\odot]$ | $0.52 \pm 0.02$ |
| $R_p[R_\odot]$ | $1.00 \pm 0.02$ | $R_p[R_\odot]$ | $1.00 \pm 0.02$ |
| $R_s[R_\odot]$ | $0.78 \pm 0.02$ | $R_s[R_\odot]$ | $0.78 \pm 0.02$ |
| $\log g_p$ | $4.40 \pm 0.02$ | $\log g_p$ | $4.40 \pm 0.02$ |
| $\log g_s$ | $4.37 \pm 0.02$ | $\log g_s$ | $4.37 \pm 0.02$ |
| $M_{bol}^p$ | $4.98 \pm 0.02$ | $M_{bol}^p$ | $4.98 \pm 0.02$ |
| $M_{bol}^s$ | $5.42 \pm 0.03$ | $M_{bol}^s$ | $5.44 \pm 0.03$ |
| $a_{orb}[R_\odot]$ | $2.30 \pm 0.02$ | $a_{orb}[R_\odot]$ | $2.30 \pm 0.02$ |

$n$ - total number of the B, V, R and I-band observations; $\Sigma(O-C)^2$ - final sum of squares of residuals between observed (LCO) and synthetic (LCC) light-curves; $\sigma_{rms}$ - root-mean-square of the residuals; $q=m_s/m_p$ - mass ratio of the components; $T_{p,s}$ - temperature of the more massive primary and less massive secondary component; $\beta_{p,s}$, $A_{p,s}$, $f_{p,s}$ - gravity-darkening exponents, albedos, and non-synchronous rotation coefficients of the components, respectively; $A_{cs}$, $\theta_{cs}$, $\lambda_{cs}$ i $\varphi_{cs}$ - the temperature coefficient, angular dimension, longitude, and latitude (in arc degrees) of the cool spot; $F_p$ - filling factor for the critical Roche lobe of the more massive, primary component; $i$ [°] - orbital inclination (in arc degrees); $l_3/(l_1+l_2+l_3)_{B,V,R,I}$ - third light contribution to the total light of the system in the B, V, R, and I band, $\Omega_{p,s}$, $\Omega_{in}$, $\Omega_{out}$ - dimensionless surface potentials of the components, and of the inner and outer contact surfaces, respectively; $f_{over}[\%]$ - degree of over-contact; $R_{p,s}$ - polar radii of the components in units of the distance between their centers; $L_p/(L_p+L_s)$ - luminosity (B; V; R; I) of the more massive, primary component (including the cool spot); $m_{p,s}[m_\odot]$, $R_{p,s}[R_\odot]$, - stellar masses, and mean radii of the components in solar units; $\log g_{p,s}$ - logarithm (base 10) of the effective gravitational acceleration of the components in the CGS units; $M_{bol}^{p,s}$ - absolute bolometric magnitudes of the components; $a_{orb}[R_\odot]$ - orbital semi-major axis in units of solar radius.

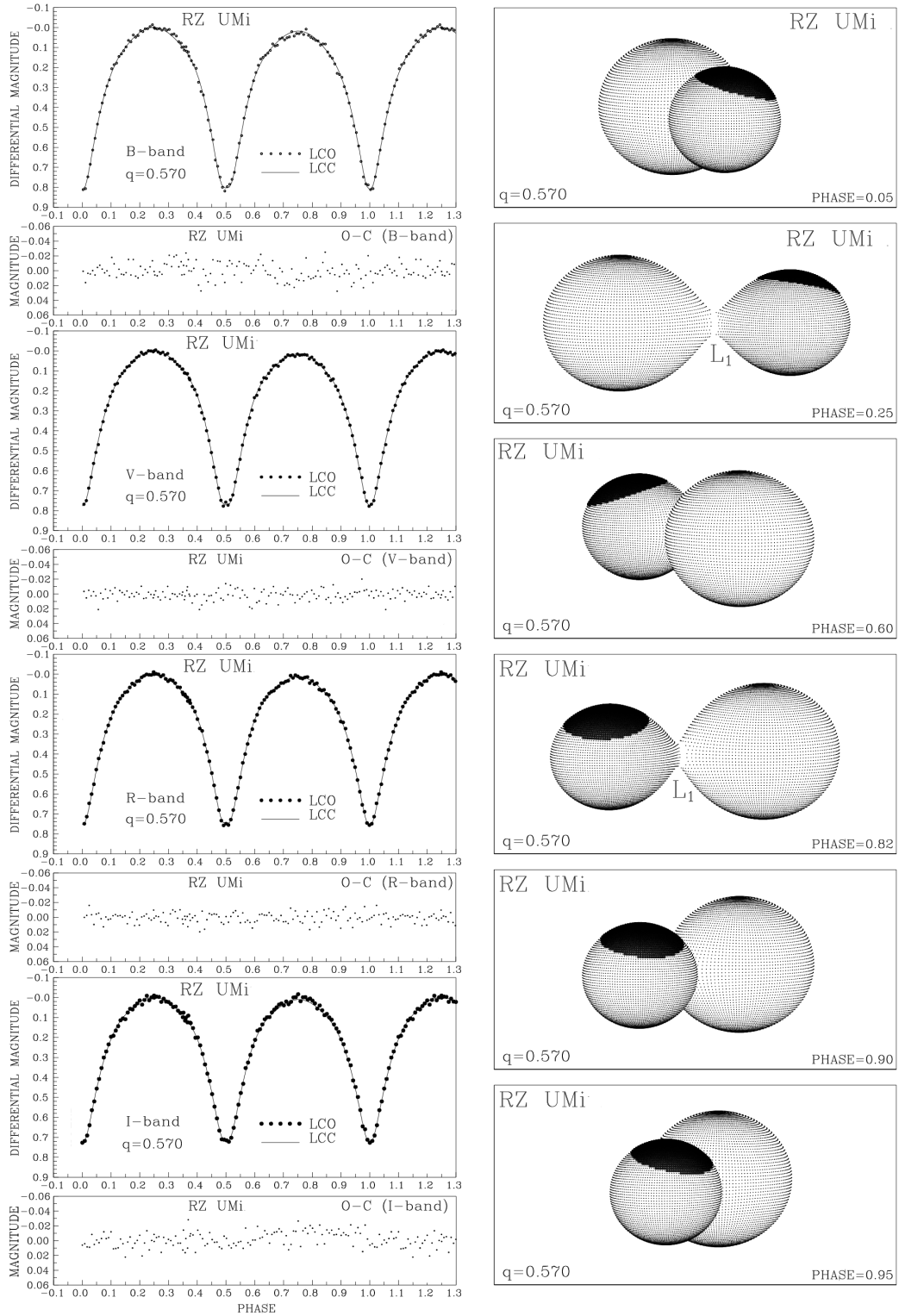

**Fig. 2**: The results of the analysis of the B, V, R, and I light curves (LCO), presented together with the optimal synthetic curve (LCC), the final $O-C$ residuals, and the model visualization with a cool spot on the secondary component at orbital phases 0.05, 0.25, 0.60, 0.82, 0.90, and 0.95.

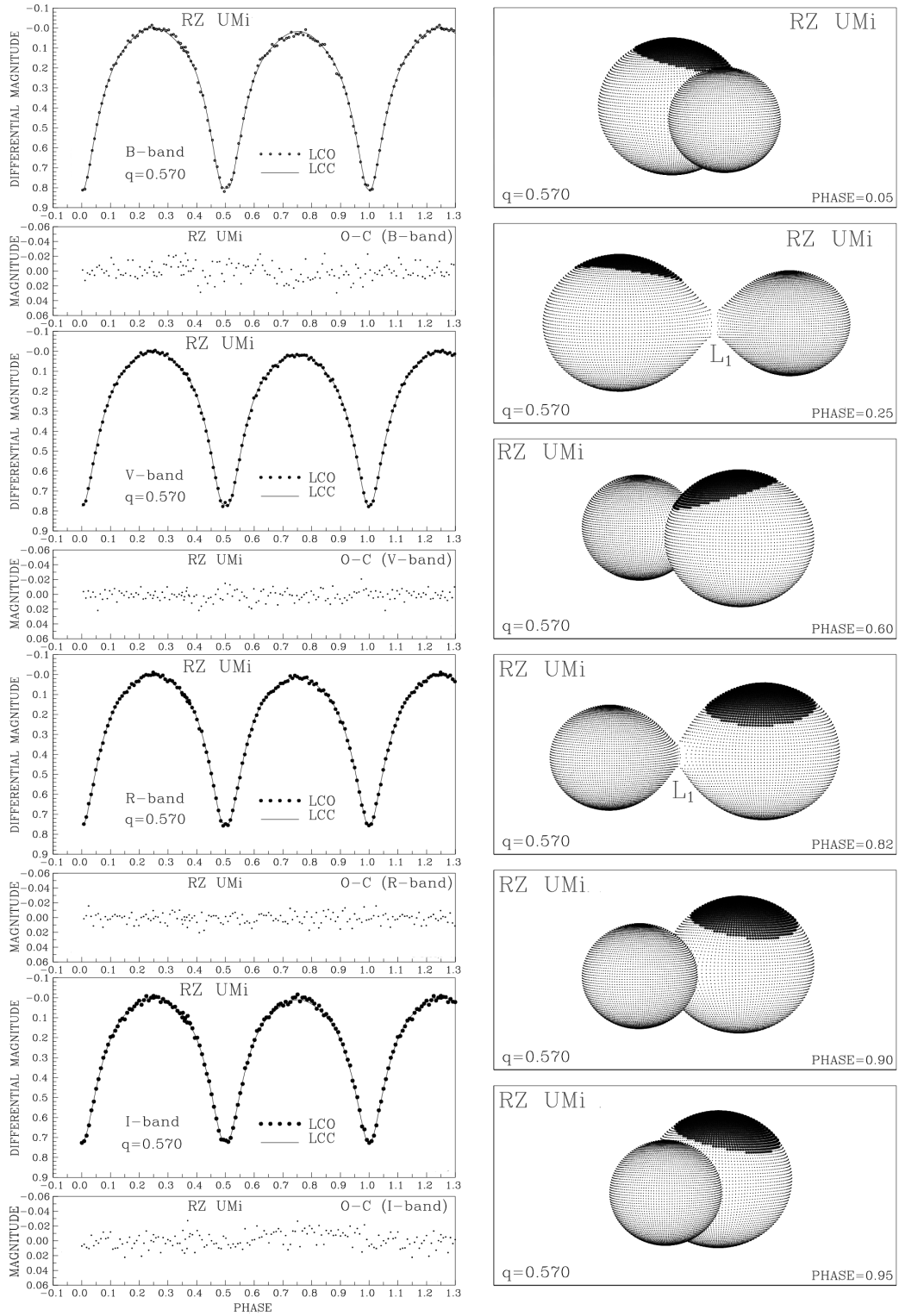

**Fig. 3**: The results of the analysis of the B, V, R, and I light curves (LCO), presented together with the optimal synthetic curve (LCC), the final $O-C$ residuals, and the model visualization with a cool spot on the primary component at orbital phases 0.05, 0.25, 0.60, 0.82, 0.90, and 0.95.

Table 4: A part of the light curve data around the primary minimum for the OQ UMa system in the I filter

| HJD | Phase | $\Delta m$ | Error |
|---|---|---|---|
| 2460409.2979 | 0.9564 | 1.3910 | 0.0062 |
| 2460409.3007 | 0.9664 | 1.4036 | 0.0045 |
| 2460409.3035 | 0.9761 | 1.4128 | 0.0045 |
| 2460409.3062 | 0.9859 | 1.4271 | 0.0046 |
| 2460409.3090 | 0.9956 | 1.4277 | 0.0046 |
| 2460409.3118 | 0.0053 | 1.4279 | 0.0047 |
| 2460409.3145 | 0.0151 | 1.4292 | 0.0049 |
| 2460409.3173 | 0.0248 | 1.4155 | 0.0049 |
| 2460409.3200 | 0.0345 | 1.4032 | 0.0047 |
| 2460409.3228 | 0.0443 | 1.3596 | 0.0044 |
| 2460409.3256 | 0.0540 | 1.3127 | 0.0045 |
| 2460409.3283 | 0.0638 | 1.2689 | 0.0045 |

Using the Kwee and van Woerden (1956) method, the times of primary and secondary minima were calculated as follows:

$$I_{min} = 2460409^d.3102 \pm 0.0014,$$

$$II_{min} = 2460409^d.4566 \pm 0.0012,$$

thus, the new ephemeris is:

$$I_{min}[HJD] = 2460409^d.3102 + 0.2833 \times E.$$

The analysis begins by determining the mass ratio of the components using the $q$-search method due to the lack of spectroscopic observations. The $q$-search for this system gives a value of $q = 0.290$ (Fig. 4). With this mass ratio, the system's inclination is found to be 89.4°. Due to such a high inclination, total eclipses occur, making the mass ratio reliable. The light curve during the secondary eclipse shows a flat-bottomed minimum because the larger component completely obscures the secondary, which has nearly half the radius and one third the mass of the primary component (Fig. 5). The light curve analysis indicates that this system is in a shallow overcontact with contact degree of $f_{over} = 12.9\%$. In the primary minimum, the larger, more massive primary component is eclipsed, classifying this system as an A-subtype. The temperature of the primary component was taken from Gaia DR3 and used as a fixed parameter. The temperature of the secondary component was found to be identical, with an uncertainty of about 35 K. The light curve exhibits different amplitudes at brightness maxima and noticeable asymmetry, suggesting the presence of a cool spot on the primary component and a bright spot on the neck region of the secondary. These features, along with the nearly equal component temperatures, are consistent with active mass transfer from the more massive to the less massive component. In addition, the detection of the third light in the I filter may indicate the presence of a cooler third body in the system (Liu et al. 2014). The analysis confirms the existence of a spot on the larger component at a latitude of 182° and a longitude of 33°, suggesting magnetic activity on the star's surface. The deviations between the observed and synthetic curves are small, indicating a good fit quality (Fig. 5). The obtained parameters are presented in Table 5.

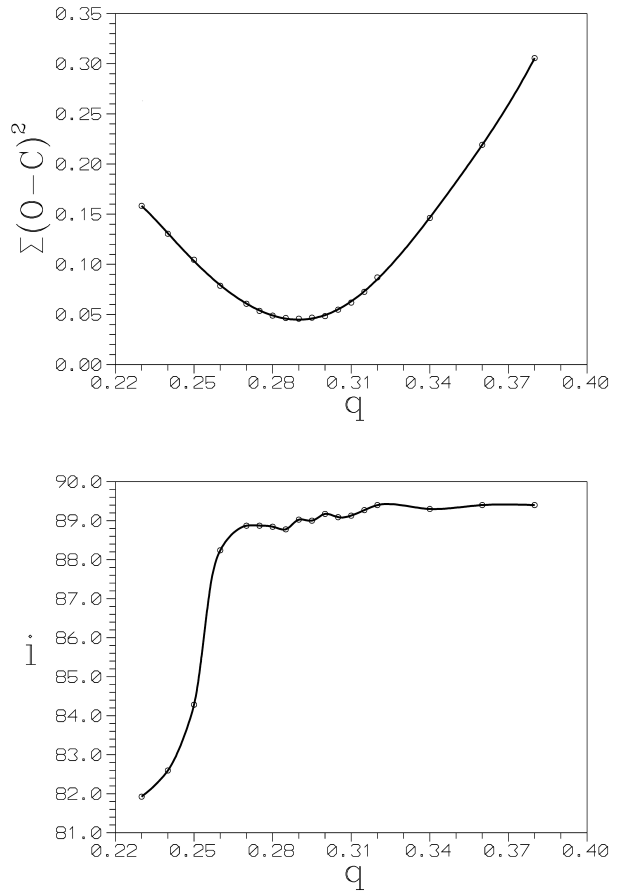

Fig. 4: Searching for the mass ratio and inclination using the $q$-search method. Each point on the upper graph represents the quality of the light curve fit, $\sum_i (O - C)^2$ for the assumed mass ratio of the components.

### 3.3. LP UMa

LP UMa was discovered as a $\delta$ Scuti-type variable star (Martin 2000). Later, observations in the V and R filters, based on the color index, estimated its spectral type to be mid-G, which is inconsistent with the classification of the $\delta$ Scuti-type stars, which are usually of spectral type A0 to F5. The light curve exhibits varying depths of minima that repeat cyclically, leading to conclusion that LP UMa is a $\beta$ Lyrae type semi-contact binary (Biro 2000) with a high mass ratio of $q = 0.88$.

According to other authors, the effective temperature was estimated based on the color index to be $T = 5500$ K with temperature difference $\Delta T = T_1 - T_2 = $ 1045 K and a contact degree $f = 57\%$ (Csizmadia et al. 2003). The light curve analysis from previous studies showed slight asymmetries in the primary and secondary minima (the O'Connell effect), which were explained by presence of a dark spot on the secondary component. The orbital period was also analyzed, and the results indicated an extremely high rate of period increase, which could be explained by mass transfer from the less massive to the more massive component, and the presence of a third body in the system.

**Table 5**: The results of simultaneous analysis of the BVRI light curves of the close binary star OQ UMa, obtained by solving the inverse problem for the Roche model with a cool spot on the more massive primary component and a bright spot in the neck region of the less massive secondary component.

| Quantity | |
|---|---|
| system | OQ UMa |
| $n(B + V + R + I)$ | 586 |
| $\Sigma(O - C)^2$ | 0.0259 |
| $\sigma_{rms}$ | 0.0067 |
| $q = m_s/m_p$ | 0.290 |
| $T_p$ | 5850 |
| $A_{p,s}$ | 0.5 |
| $\beta_{p,s}$ | 0.08 |
| $f_p = f_s$ | 1.0 |
| $A_{cs} = T_{cs}/T_s$ | $0.95 \pm 0.02$ |
| $\theta_{cs}$ | $32.9 \pm 1.5$ |
| $\lambda_{cs}$ | $182.5 \pm 2.6$ |
| $\varphi_{cs}$ | $33.1 \pm 1.0$ |
| $A_{bs} = T_{bs}/T_s$ | $1.10 \pm 0.03$ |
| $\theta_{bs}$ | $43.4 \pm 1.8$ |
| $\lambda_{bs}$ | $155.5 \pm 2.6$ |
| $\varphi_{bs}$ | $37.6 \pm 1.0$ |
| $T_s$ | $5849 \pm 35$ |
| $F_p$ | $1.011 \pm 0.002$ |
| $i\ [°]$ | $89.4 \pm 0.4$ |
| $l_3/(l_1 + l_2 + l_3)_B$ | $0.000 \pm 0.002$ |
| $l_3/(l_1 + l_2 + l_3)_V$ | $0.008 \pm 0.002$ |
| $l_3/(l_1 + l_2 + l_3)_R$ | $0.004 \pm 0.002$ |
| $l_3/(l_1 + l_2 + l_3)_I$ | $0.035 \pm 0.002$ |
| $\Omega_{p,s}$ | 2.4206 |
| $\Omega_{in}$ | 2.4440 |
| $\Omega_{out}$ | 2.2628 |
| $f_{over}[\%]$ | 12.9 |
| $R_p[D = 1]$ | 0.463 |
| $R_s[D = 1]$ | 0.264 |
| $L_p/(L_p + L_s)(B; V; R; I)$ | 0.729; 0.731; 0.732; 0.733 |
| $m_p[M_\odot]$ | $1.00 \pm 0.02$ |
| $m_s[M_\odot]$ | $0.29 \pm 0.02$ |
| $R_p[R_\odot]$ | $0.98 \pm 0.02$ |
| $R_s[R_\odot]$ | $0.56 \pm 0.02$ |
| $\log g_p$ | $4.45 \pm 0.02$ |
| $\log g_s$ | $4.40 \pm 0.02$ |
| $M_{bol}^p$ | $4.77 \pm 0.02$ |
| $M_{bol}^s$ | $5.98 \pm 0.03$ |
| $a_{orb}[R_\odot]$ | $1.97 \pm 0.02$ |

Prasad et al. (2014) published new photometric results in a study of three W UMa-type contact systems. They found a small O'Connell effect due to

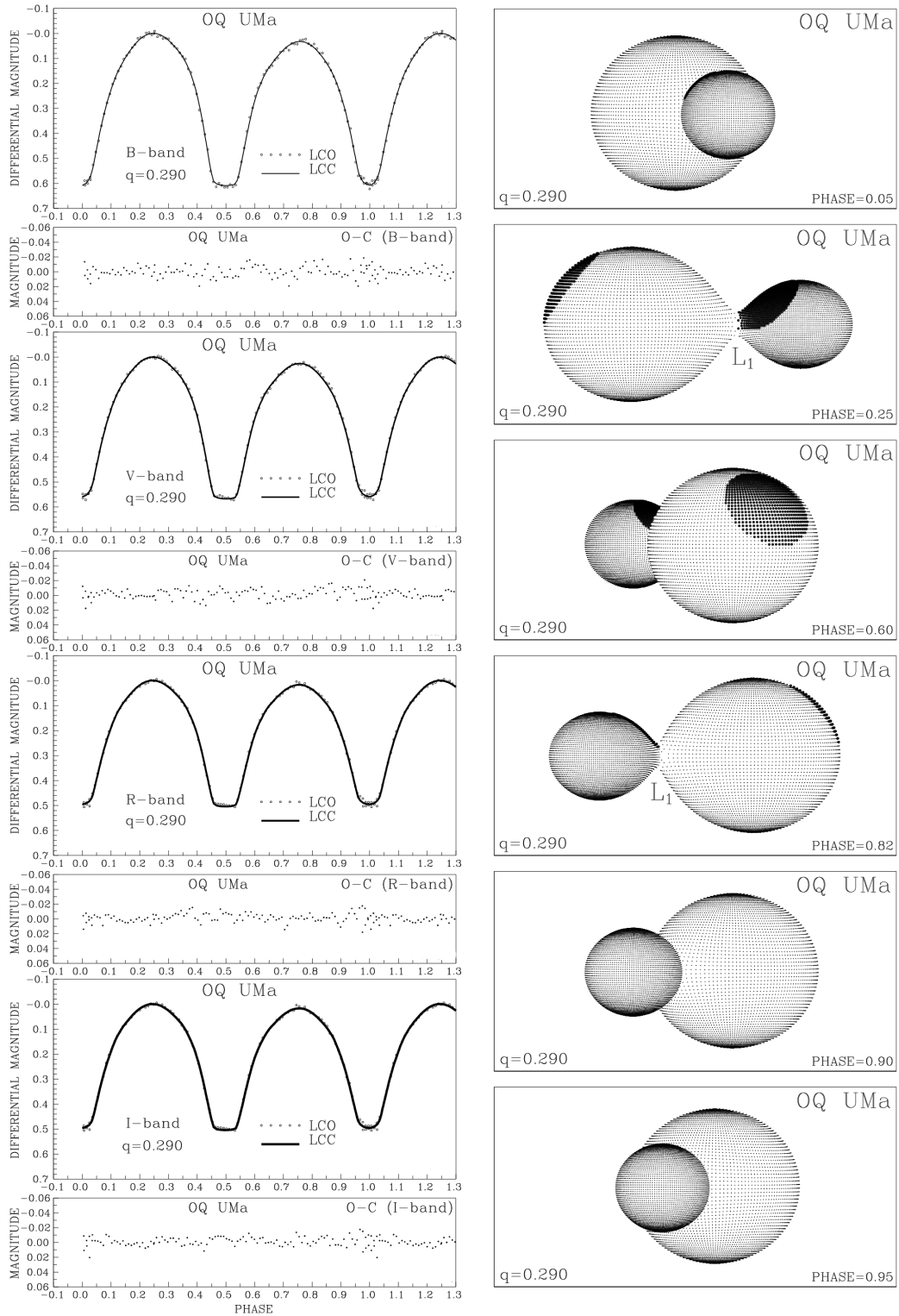

**Fig. 5**: The results of analysis of the B, V, R, and I light curves (LCO), presented together with the optimal synthetic curve (LCC), the final O-C residuals, and the model visualization featuring a cool spot on the primary component and a bright spot in the neck region of the secondary component at orbital phases 0.05, 0.25, 0.60, 0.82, 0.90, and 0.95.

43

presence of a hot spot on the components. Additionally, they determined that LP UMa is a contact binary with a contact degree of $f = 14\%$ and a high mass ratio, consistent with Csizmadia et al. (2003). The change in orbital period was attributed to mass transfer from the primary to the secondary component.

A new set of light curves and eight new eclipse timings were presented in the study by Liao et al. (2015). They confirmed the fastest period increase rate for the W UMa-type stars ($+10.21\,\mathrm{s}/100\,\mathrm{yr}$), suggesting that the primary star is accreting mass from the secondary star at a rate of $\dot{m} = 5.3 \times 10^{-6}\,M_\odot\mathrm{yr}^{-1}$. With the inclusion of a third component, the absolute physical parameters were estimated as $m_1 = 0.9\,M_\odot$, $m_2 = 0.74\,M_\odot$, $R_1 = 1.04\,R_\odot$, $R_2 = 0.97\,R_\odot$, $L_1 = 0.875\,L_\odot$, $L_2 = 0.442\,L_\odot$, separation $a = 2.27\,R_\odot$, mass ratio $q = 0.823$, and contact degree $f = 66.6\%$. Additional spectroscopic observations are needed to confirm the mass ratio and investigate whether an additional third component contributes to the total luminosity and plays a significant role in the system's evolution.

Observations and analyses from the Weihai Observatory yielded significantly different results (Guo et al. 2016). Their study concluded that LP UMa is an A-subtype W UMa-type contact binary with a mass ratio $q = 0.33$, temperature difference between components $\Delta T = 90\,\mathrm{K}$, and contact degree of $f = 7.9\%$. The asymmetric light curve was also explained by presence of a hot spot on the more massive component. Given that both components are the main-sequence stars, primary component's mass was estimated at $M_1 = 0.92\,M_\odot$ and the secondary at $M_2 = 0.30\,M_\odot$. The orbital period also showed a high rate of increase, consistent with previous studies, but the mass accretion rate explaining this change differed by an order of magnitude ($\dot{m} = 5.2 \times 10^{-7}\,M_\odot yr^{-1}$), likely due to the different mass ratios.

With new photometric data, we will conduct another analysis of this system and compare the obtained results. As in previous cases, we first determine the epoch, which begins with the deeper minimum, which in this case is also the primary minimum. Using the Kwee and van Woerden (1956) method, the times of the primary and secondary minima were determined as:

$$I_{\min} = 2460340^d.6245 \pm 0.0003,$$
$$II_{\min} = 2460340^d.4733 \pm 0.0003,$$

thus, the new ephemeris is:

$$I_{\min}[\mathrm{HJD}] = 2460340^d.6245 + 0.3098 \times E.$$

A portion of the observational data used to construct the light curve is given in Table 6.

When comparing the light curves obtained in previous studies, it can be observed that their shape has changed over time. The depths of the minima have nearly equalized, revealing that the effective temperatures of both components are becoming similar, as indicated by the results of the analysis. Due to the asymmetry of the light curve, we analyzed the data under the hypothesis that there is a cool spot on the surface of both the primary and secondary components, with their coordinates determined by the program.

Table 6: A part of the light curve data around the primary minimum for the LP UMa system in the R filter

| HJD | Phase | $\Delta m$ | Error |
|---|---|---|---|
| 2460340.6204 | 0.9870 | -0.0948 | 0.0037 |
| 2460340.6215 | 0.9904 | -0.0987 | 0.0037 |
| 2460340.6226 | 0.9940 | -0.0999 | 0.0037 |
| 2460340.6237 | 0.9974 | -0.0967 | 0.0037 |
| 2460340.6248 | 0.0009 | -0.0990 | 0.0037 |
| 2460340.6259 | 0.0044 | -0.1026 | 0.0037 |
| 2460340.6269 | 0.0079 | -0.0929 | 0.0037 |
| 2460340.6280 | 0.0114 | -0.1011 | 0.0037 |
| 2460340.6291 | 0.0149 | -0.1064 | 0.0037 |
| 2460340.6302 | 0.0183 | -0.1055 | 0.0037 |
| 2460340.6313 | 0.0218 | -0.1043 | 0.0037 |
| 2460340.6323 | 0.0254 | -0.1018 | 0.0037 |

During the parameter determination process, the temperature of the primary component was fixed at previously determined value of $T_p = 5500\,\mathrm{K}$ (Csizmadia et al. (2003), Liao et al. (2015)). The effects of gravity darkening were also fixed at $g = 0.32$ for convective atmospheres (Lucy 1968), as well as the values for bolometric albedo at $g = 0.32$ (Ruciński 1969). The parameters set as free variables included the inclination, temperature of the secondary component, monochromatic luminosity of the primary component, and dimensionless potentials for both stars. The mass ratio was taken from the work Guo et al. (2016), where the authors conducted a $q$-search and obtained a value of $q = 0.327$.

The analysis data indicate that the LP UMa system belongs to the A subtype of the W UMa class, with a contact degree of $f = 17.12\%$, and has an active atmosphere with dark spots. Considering that these are the main-sequence stars, the mass of the primary component is estimated to be $m_p = 0.921\,m_\odot$ according to Cox (2000), and from the mass ratio, the less massive component has a mass of $m_s = 0.301\,m_\odot$. System's inclination is determined to be $i = 50.3°$, indicating that total eclipses do not occur, so the derived mass values may vary, and these results should be considered preliminary. Spectroscopic observations are required for this system to confirm the mass ratio, which would allow for more precise calculations of other parameters. The obtained parameters are presented in Table 7.

In Fig. 6, the left-hand side shows the light curves through the observed filters, where the points represent the observed values, while the solid line fitting

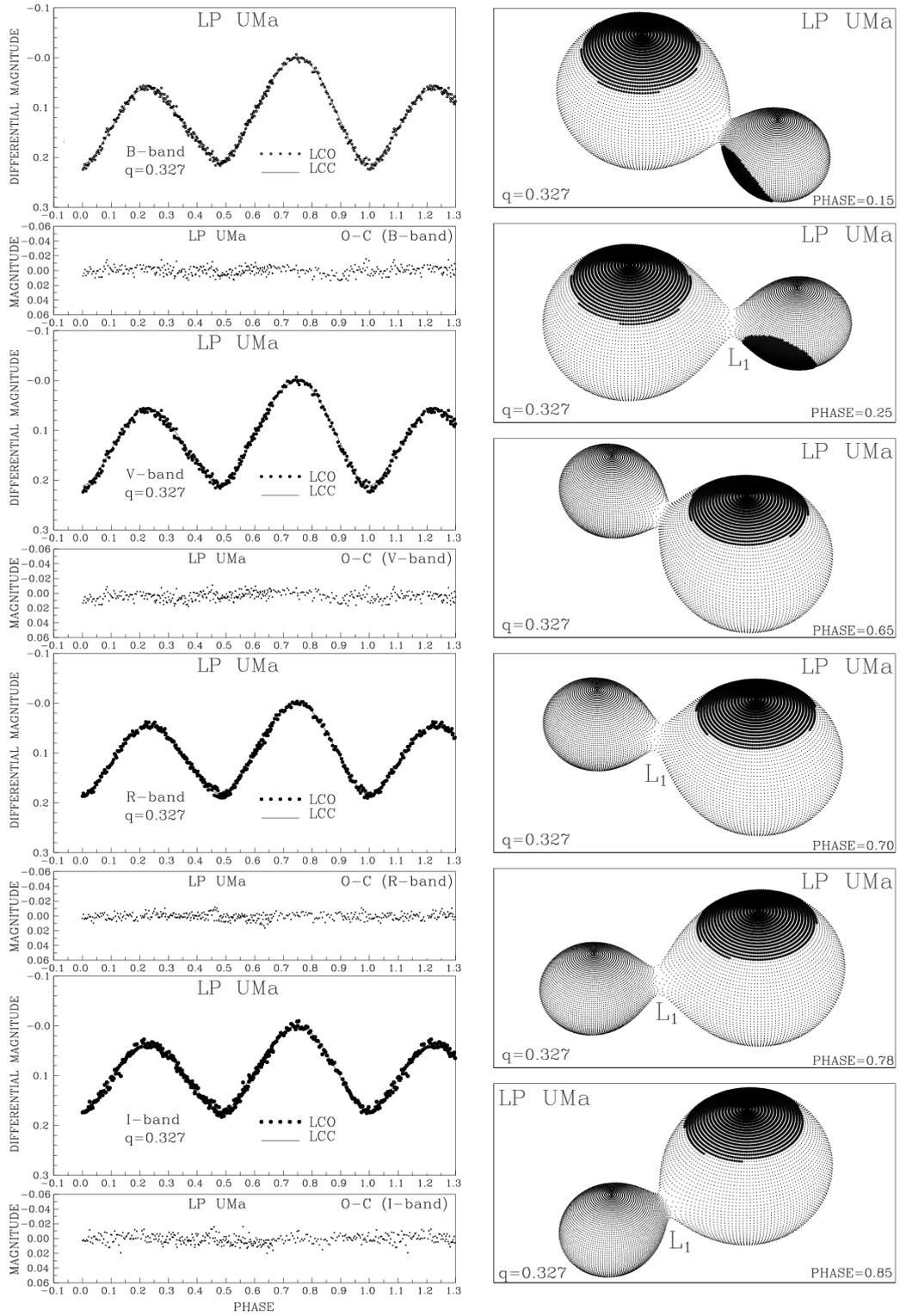

**Fig. 6**: The results of analysis of the B, V, R, and I light curves (LCO), presented together with the optimal synthetic curve (LCC), the final O-C residuals, and the model visualization with a cool spot on both the primary and secondary components at orbital phases 0.15, 0.25, 0.65, 0.70, 0.78, and 0.85.

45

Table 7: The results of simultaneous analysis of the BVRI light curves of the binary star LP UMa, obtained by solving the inverse problem for the Roche model with a cool spot on both the primary and secondary components.

| Quantity | |
|---|---|
| system | LP UMa |
| $n(B + V + R + I)$ | 1585 |
| $\Sigma(O - C)^2$ | 0.0473 |
| $\sigma_{\mathrm{rms}}$ | 0.0054 |
| $q = m_{\mathrm{s}}/m_{\mathrm{p}}$ | 0.327 |
| $T_{\mathrm{p}}$ | 5500 |
| $A_{\mathrm{p,s}}$ | 0.5 |
| $\beta_{\mathrm{p,s}}$ | 0.08 |
| $f_{\mathrm{p}} = f_s$ | 1.0 |
| $A_{\mathrm{cs}} = T_{\mathrm{cs}}/T_{\mathrm{p}}$ | $0.85 \pm 0.02$ |
| $\theta_{\mathrm{cs}}$ | $48.1 \pm 1.0$ |
| $\lambda_{\mathrm{cs}}$ | $69.1 \pm 2.0$ |
| $\varphi_{\mathrm{cs}}$ | $86.3 \pm 1.0$ |
| $A_{\mathrm{cs}} = T_{\mathrm{cs}}/T_{\mathrm{s}}$ | $0.87 \pm 0.02$ |
| $\theta_{\mathrm{bs}}$ | $46.5 \pm 1.5$ |
| $\lambda_{\mathrm{bs}}$ | $117.6 \pm 2.0$ |
| $\varphi_{\mathrm{bs}}$ | $-29.3 \pm 1.0$ |
| $T_{\mathrm{s}}$ | $5512 \pm 30$ |
| $F_{\mathrm{p}}$ | $1.015 \pm 0.002$ |
| $i\,[°]$ | $50.3 \pm 0.3$ |
| $l_3/(l_1 + l_2 + l_3)_B$ | $0.0060 \pm 0.003$ |
| $l_3/(l_1 + l_2 + l_3)_V$ | $0.000 \pm 0.002$ |
| $l_3/(l_1 + l_2 + l_3)_R$ | $0.077 \pm 0.003$ |
| $l_3/(l_1 + l_2 + l_3)_I$ | $0.087 \pm 0.002$ |
| $\Omega_{\mathrm{p,s}}$ | 2.4905 |
| $\Omega_{\mathrm{in}}$ | 2.5252 |
| $\Omega_{\mathrm{out}}$ | 2.3225 |
| $f_{\mathrm{over}}[\%]$ | 17.12 |
| $R_{\mathrm{p}}[D=1]$ | 0.456 |
| $R_{\mathrm{s}}[D=1]$ | 0.275 |
| $L_{\mathrm{p}}/(L_{\mathrm{p}}+L_{\mathrm{s}})(B;V;R;I)$ | 0.721; 0.722; 0.722; 0.723 |
| $m_{\mathrm{p}}[M_\odot]$ | $0.92 \pm 0.02$ |
| $m_{\mathrm{s}}[M_\odot]$ | $0.30 \pm 0.02$ |
| $R_{\mathrm{p}}[R_\odot]$ | $1.01 \pm 0.02$ |
| $R_{\mathrm{s}}[R_\odot]$ | $0.61 \pm 0.02$ |
| $\log g_{\mathrm{p}}$ | $4.44 \pm 0.02$ |
| $\log g_{\mathrm{s}}$ | $4.34 \pm 0.02$ |
| $M_{\mathrm{bol}}^{\mathrm{p}}$ | $4.98 \pm 0.02$ |
| $M_{\mathrm{bol}}^{\mathrm{s}}$ | $6.05 \pm 0.03$ |
| $a_{\mathrm{orb}}[R_\odot]$ | $2.06 \pm 0.02$ |

the points represents the synthetic values the parameters were derived from. Below the curve, the $O - C$ indicates the quality of the fit. On the right-hand side, a 3D model of the system is displayed at different orbital phases from the observer's perspective, with an inclination of 50°. The images confirm that total eclipses do not occur, which is also evident in the light curves due to the small amplitude of minima.

## 4. DISCUSION AND CONCLUSION

In this work, an extensive photometric analysis of the CB systems RZ UMi, OQ UMa, and LP UMa was carried out, and their geometric and physical parameters were established. New times of the primary and secondary minima were derived, which can be utilized in the future period change studies that may be attributed to the ongoing mass transfer in the systems. All three CB exhibit light curve variability and shallow contact configurations with contact degrees from 9% to 17%.

For the temperature inputs of the primary components, we adopted values based on previously published studies. For RZ UMi, van Cauteren et al. (2006) determined the primary temperature using the period–color relation given by Eggen (1967), and the adopted value is in close agreement with the Gaia DR3 effective temperature. For LP UMa, previous studies, including Csizmadia et al. (2003), Guo et al. (2016), Prasad et al. (2014), Liao et al. (2015), consistently reported a primary temperature around 5500 K, which is in close agreement with the value provided by Gaia DR3. Therefore, we adopted this commonly used value for the LP UMa. In the case of OQ UMa, no earlier temperature estimates were available in the literature; therefore, the effective temperature of the primary component was directly taken from Gaia DR3. In all cases, the adopted temperatures are consistent with Gaia DR3 values and provided a stable convergence during the light-curve modeling.

RZ UMi and OQ UMa both exhibit total eclipses due to their high inclinations, which strongly constrain the mass ratios. Compared to the earlier attempts of van Cauteren et al. (2006) for RZ UMi, when they gave a wide set of possible mass ratios and quite uncertain parameters, our multiband photometric solution provides a more limited parameters and physically acceptable solution. The $q$-search method for mass ratio estimation, performed photometrically in the absence of spectroscopic data, yielded well-defined minima of the fitting quality function, providing high confidence in the derived mass ratios. The secondary star in RZ UMi was found to be slightly hotter than their more massive primary star, despite the fact that the system belongs to the A subtype, which is a rare occurrence (Alton and Stępień 2021). However, this classification should be taken with caution due to small differences in temperature.

For the case of OQ UMa, our solution is the first detailed light curve modeling and shows some discrepancies with the parameters provided by Khalatyan et al. (2024), likely caused by different modeling approaches. Furthermore, a third light was significantly detected in the I filter but was not significant in the B, V, and R bands. The results reveal that there is also a third light in the LP UMa system, as shown in Table 7, the third light contribution is significant in the R and I band. This wavelength-dependent characteristic can imply the presence of a redder tertiary component. Similar phenomena have been observed in other contact binaries where cooler third components become apparent only in longer wavelength bands (Liu et al. 2014).

The asymmetries in the light curves in all of the systems were appropriately modeled using star spots. For RZ UMi the best fit was with a cool spot on either the primary or secondary star, both of which yielded comparable solutions. OQ UMa required a cool spot on the primary and a bright spot on the secondary's neck to match the observed O'Connell effect and to explain the nearly identical component temperatures, which may be a sign of ongoing mass transfer. In addition to determining the fundamental parameters of the system, the low values of residuals ($\sigma_{\rm rms} \sim 0.005 - 0.009$) and the minimized values of $\Sigma(\rm O - C)^2$ for all systems demonstrate the reliability of the inverse problem method based on the Roche model.

LP UMa, in contrast, is a low-inclination system that produces partial eclipses, limiting the accuracy of photometric mass ratio determination. Thus, we adopted the previously calculated value of Guo et al. (2016). Our solutions are generally in agreement with previous studies in classification as an overcontact binary, mass transfer activity, and presence of surface spots. However, there are considerable discrepancies in the adopted mass ratio and contact degree (e.g. Csizmadia et al. (2003), Liao et al. (2015), suggesting a higher contact degree and larger temperature differences). We found a much lower temperature difference, and a moderate contact degree ($f = 17.12\%$). This supports the commonly observed characteristic in contact binary systems that their components have nearly equal temperatures (Lucy 1968). To verify the exact configuration for LP UMa, a spectroscopically determined mass ratio is required as input for the photometric light curve modeling, but such observations are not yet available. Consequently, the absolute parameters of the LP UMa system remain uncertain, and the presented solution should be considered with caution. Overall, this work demonstrates that detailed multiband photometric modeling, when combined with appropriate physical assumptions and spot modeling, can yield system parameters even in absence of spectroscopic information.

*Acknowledgements* – Authors acknowledge the support from the Ministry of Science, Technological Development and Innovation of the Republic of Serbia (MSTDIRS) through contract no. 451-03-136/2025-03/200002, made with the Astronomical Observatory (Belgrade, Serbia).

## АНАЛИЗА КРИВИХ СЈАЈА ТЕСНИХ ДВОЈНИХ СИСТЕМА ТИПА W-UMa: RZ UMi, OQ UMa, LP UMa

М. Гроздановић и Г. Ђурашевић

*Астрономска опсерваторија, Волгина 7, 11060 Београд 38, Србија*

E–mail: *mgrozdanovic@aob.rs*

У овом раду представљена су нова висококвалитетна фотометријска оптичка посматрања три тесна двојна система типа W-UMa. Анализа одговарајућих кривих сјаја извршена је применом Ђурашевићевог кода за инверзни проблем. Ради објашњења асиметрија и варијација у кривама сјаја, коришћен је Рошов модел са пегама на компонентама система. Одређени су основни параметри, укључујући однос маса добијен методом $q$-претраге. Испитиване су хипотезе које подразумевају постојање активних површинских области, као што су тамне пеге на примарној и секундарној компоненти или светле пеге на регији врата, које су резултат магнетне активности или континуираног преноса масе између компоненти, како би се објасниле различите амплитуде максимума и готово једнаке дубине минимума на кривама сјаја. На основу добијених орбиталних и физичких параметара, конструисани су тродимензионални модели за различите орбиталне фазе.